\documentclass[journal=jacsat,manuscript=article]{achemso}

\usepackage{chemformula} 
\usepackage[T1]{fontenc} 

\usepackage{amsmath}
\usepackage{amstext}
\usepackage{amssymb}
\usepackage[per=frac,separate-uncertainty = true]{siunitx}
\usepackage{xargs}
\usepackage{subcaption}
\usepackage{epstopdf}

\author{Timo Weggler}
\email{jens.michaelis@uni-ulm.de}
\phone{+49 (0)731 23050}
\affiliation[Biophysics University Ulm]
{Institute for Biophysics and Center for Integrated Quantum Science and Technology (IQST), Universit\"at Ulm, Albert-Einstein-Allee 11, 89069 Ulm, Germany}
\author{Christian Ganslmayer}
\affiliation[Biophysics University Ulm]
{Institute for Biophysics, Universit\"at Ulm, Albert-Einstein-Allee 11, 89069 Ulm, Germany}
\author{Florian Frank}
\affiliation[Quantum Optics University Ulm]
{Institute for Quantum Optics, Universit\"at Ulm, Albert-Einstein-Allee 11, 89069 Ulm, Germany}
\author{Tobias Eilert}
\affiliation[Biophysics University Ulm]
{Institute for Biophysics, Universit\"at Ulm, Albert-Einstein-Allee 11, 89069 Ulm, Germany}
\author{Fedor Jelezko}
\affiliation[Quantum Optics University Ulm]
{Institute for Quantum Optics, Universit\"at Ulm, Albert-Einstein-Allee 11, 89069 Ulm, Germany}

\author{Jens Michaelis}
\affiliation[Biophysics University Ulm]
{Institute for Biophysics and Center for Integrated Quantum Science and Technology (IQST), Universit\"at Ulm, Albert-Einstein-Allee 11, 89069 Ulm, Germany}

\title[]{Determination of the 3D Magnetic Field Vector Orientation with NV Centers in Diamond}

\abbreviations{NV, ODMR, ZFS}
\keywords{NV, color centers, optical detected magnetic resonance (ODMR), magnetometer, spin sensing}

\begin{document}


\textbf{}
\begin{abstract}
	Absolute knowledge about the magnetic field orientation plays a crucial role in single spin-based quantum magnetometry and the application toward spin-based quantum computation. In this paper, we reconstruct the 3D orientation of an arbitrary static magnetic field with individual nitrogen vacancy (NV) centers in diamond. We determine the polar and the azimuthal angle of the magnetic field orientation relative to the diamond lattice. Therefore, we use information from the photoluminescence anisotropy of the NV, together with a simple pulsed Optically Detected Magnetic Resonance (ODMR) experiment. Our nanoscopic magnetic field determination is generally applicable and does not rely on special prerequisites such as strongly coupled nuclear spins or particular controllable fields.
	Hence, our presented results open up new paths for precise NMR reconstructions and the modulation of the electron-electron spin interaction in EPR measurements by specifically tailored magnetic fields.
\end{abstract}
\textbf{Keywords: }NV, color centers, optical detected magnetic resonance (ODMR), magnetometer, spin sensing

\section{Introduction}
	During the last decades, color centers in diamond became important for many applications in the fields of quantum information, sensing at the nanometer level, and in particular quantum sensing.\cite{jelezko2006single} One candidate of special interest is the Nitrogen Vacancy Center (NV), as it is well characterized and can be used without special requirements to its ambient conditions. One of the most prominent application is NV based magnetometry, as it is a promising field of research with an increasing impact not only in the quantum optics community, but also for the fabrication of nano-scale sensors\cite{bucher2019quantum}. Reasons for the great interest in NV centers for novel sensing applications are the beneficial material properties of diamond \cite{DohertyNV2013} and the remarkable features of the NV-defect, i.e. its nanometer scale, high stability and the thus resulting tremendous sensitivity in magnetometry\cite{jelezko2006single}. This allows the usage of NV-based sensors in a wide field of experimental setups like NMR\cite{devience2015nanoscale,mamin2013nanoscale} and EPR\cite{schlipf2017molecular}. As NVs do not rely on cryogenic temperatures like SQUID magnometers and can be designed in nano-scale sizes, they need less experimental overhead and can also be brought into close vicinity to the sample of interest\cite{schirhagl2014nitrogen}, thus making them ideal nanoscopic sensing devices. Since the absolute knowledge about the magnetic field is crucial for many applications\cite{le2013optical,casola2018probing}, there exists heavy interest in its directional determination. However until now, only the polar angle and the B-field amplitude have been determined\cite{balasubramanian2008nanoscale}. The information of the azimuthal angle gets lost due to the C$_{3\nu}$ symmetry of the NV. Here, we break this symmetry by using a set of NVs with different orientations to sense the same magnetic field. We perform ODMR measurements to calculate the individual polar angles for the different axes and then use an algorithm to calculate the exact vector of the magnetic field. This will pave the way for sensing protocols with a higher complexity and provide a new tool for future experiments benefiting from knowledge about the magnetic field. A further interesting application of this technique is that when it is applied in a reverse scheme, it gives evidence about an arbitrary NV orientation measured in previously characterized magnetic fields.

\section{Methods and Results}
\begin{figure}
		\centering
		\includegraphics[width=.7\linewidth]{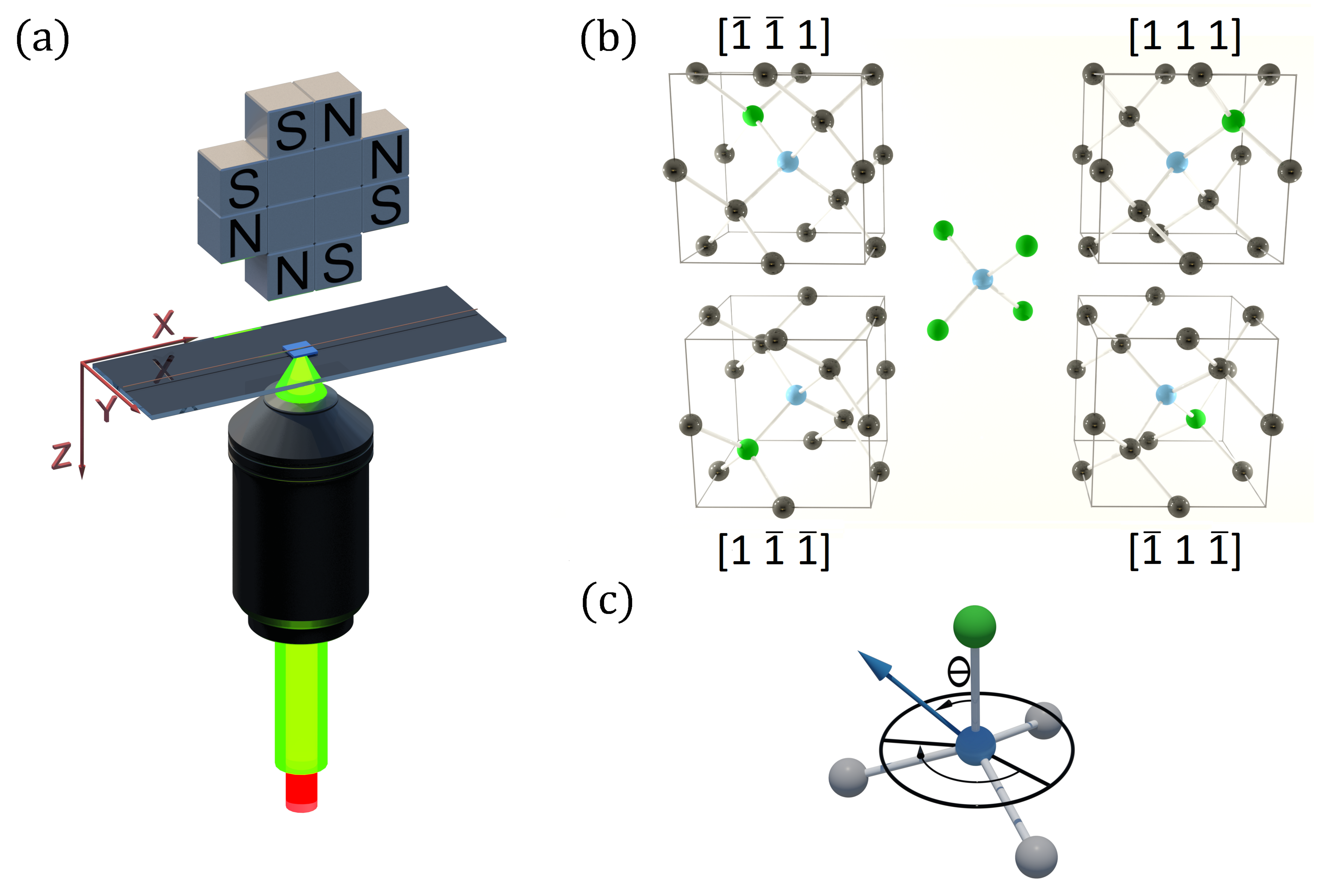}
		\caption{(a) Schematic, illustrating the experimental geometry. The diamond sample is moved between the rigidly arranged objective and magnet. (b) The four possible orientations of the principal axis of the NV center (nitrogen green; vacancy blue) combined, these orientations span a tetrahedron. (c) Orientation of the B-field (blue arrow) and the cone angle $\theta$ with respect to the NV center principal axis.}		
		\label{fig:Confocal_Head_NV_Lattice}
	\end{figure}

	In this work we present a general technique to determine the relative (or absolute, if required) orientation of the magnetic field.\\
	In our confocal microscope setup, a diamond with embedded NVs is moved between a rigidly arranged objective and magnet assembly (see fig.\ref{fig:Confocal_Head_NV_Lattice}a). In the sp$^3$-hybridized diamond lattice, the NV-center replaces two adjacent carbon atoms with a vacancy and a nitrogen atom, leading to four possible orientations with respect to the lattice (see fig.\ref{fig:Confocal_Head_NV_Lattice}b). The four directions can be reduced in a pictorial representation into one tetrahedron, with the vacancy in the middle and the nitrogen atoms at the corners (see fig.\ref{fig:Confocal_Head_NV_Lattice}b). By translation of the diamond sample, the NV-centers are moved into the center of the excitation volume. Therefore, each NV center sense the same magnetic field amplitude and absolute orientation. Due to the symmetry of the diamond structure, the angle between two of those four possible NV center directions is always $\SI{109.47}{\degree}$. To describe the magnetic field vector completely with respect to the NV centers' principal axis, i.e. the connection of the nitrogen atom and the vacancy, the tilt angle $\theta$ between the magnetic field and the NV axis is insufficient (see fig.\ref{fig:Confocal_Head_NV_Lattice}c). Hence, the polar angle $\psi$, as well as the azimuthal angle $\phi$ of the spherical coordinates need to be determined to describe the magnetic field vector absolutely in the three dimensional space.
	
\begin{figure}
		\centering
		\includegraphics[width=.5\linewidth]{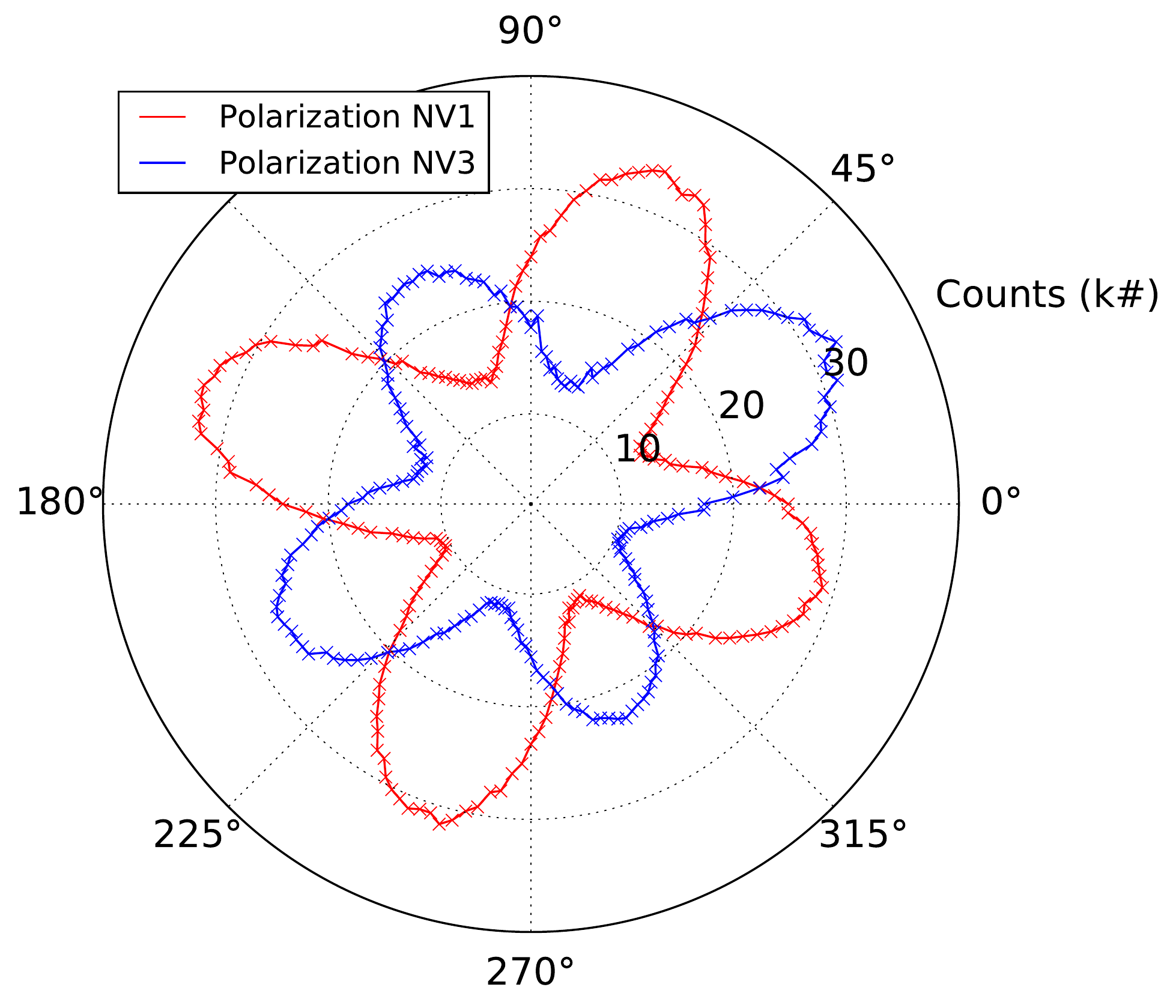}
		\caption{Photoluminescence anisotropy measurement. Rotation of the $\lambda\slash 2$-plate in the illumination pathway and the resulting fluorescence signal ($10^3$ counts per second) obtained for the two different NV axis orientations in the diamond lattice. Due to the $(100)$ cut of the diamond, the polarization axes are tilted by $\SI{90}{\degree}$, corresponding to a $\lambda\slash 2$-plate rotation angle of $\SI{45}{\degree}$.}		
		\label{fig:Polarization_Axes}
	\end{figure}

	For a diamond cut along the $(100)$ direction, the photoluminescence anisotropy of the NV-center\cite{alegre2007polarization} allows to differentiate the four possible NV directions into two pairs of orthogonal polarization(see fig.\ref{fig:Polarization_Axes}). Furthermore, for each of these pairs the orientation of the two respective NV centers can be distinguished by applying a static magnetic field parallel to the orientation of one of the two NV centers. This can be done by measuring the symmetry of the $\left|{0}\right>\rightarrow\left|{\pm1}\right>$ transition frequencies with respect to the Zero Field Splitting (ZFS). With this alignment along the principal axes, the signal of the aligned color center stays constant and the signal of the second NV-center decreases. 
	
\begin{figure}
\centering
\includegraphics[width=.8\linewidth,keepaspectratio]{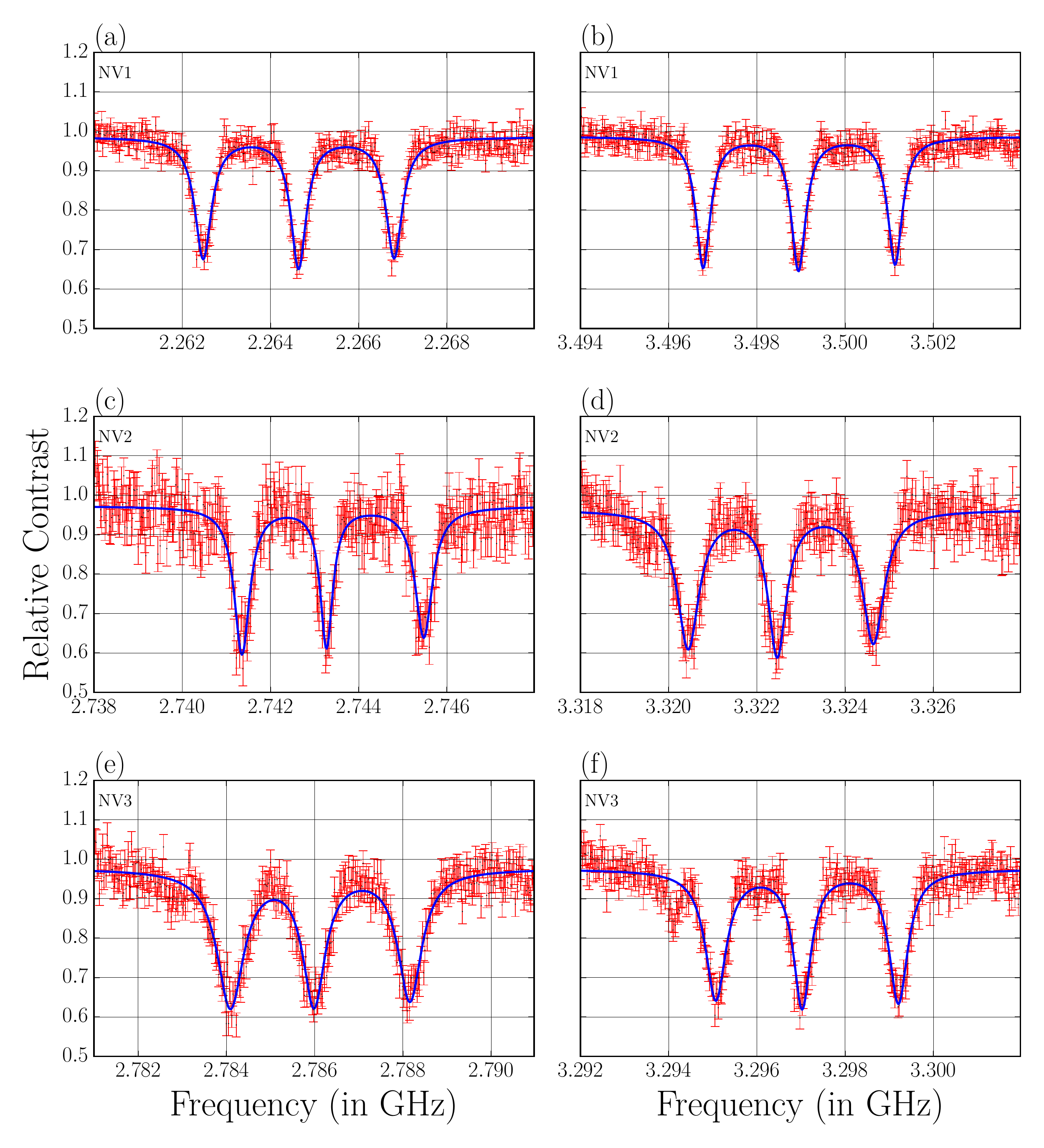}
\caption{Results for the pulsed ODMR measurement normalized to the maximum Rabi contrast. Measurements for three different NV orientations with respect to the B-field are shown. The resulting angle $\theta_{\mathrm{i}}$ is the tilt between the NV axis and the B-field orientation. For NV1, the transition frequencies $\left|0\right>\rightarrow\left|-1\right>$ (a) and $\left|0\right>\rightarrow\left|\pm1\right>$ (b) with resulting angle $\theta_{\mathrm{1}} = \SI{13.42}{\degree}$. For NV2, the transition frequencies $\left|0\right>\rightarrow\left|-1\right>$ (c) and $\left|0\right>\rightarrow\left|\pm1\right>$ (d) with resulting angle $\theta_{\mathrm{2}} = \SI{62.89}{\degree}$. For NV3, the transition frequencies $\left|0\right>\rightarrow\left|-1\right>$ (e) and $\left|0\right>\rightarrow\left|\pm1\right>$ (f) with resulting angle $\theta_{\mathrm{3}} = \SI{66.61}{\degree}$. During the measurements the NVs are driven with a Rabi period of $\Omega = \SI{1.58}{\micro\second}$ while subject to a B-field with amplitude $\left|B\right|\sim\SI{230}{Gauss}$.}
\label{fig:pulsedODMR_meas}
\end{figure}

Since the natural abundance of $N^{14}$ nitrogen atoms is $\SI{99.634}{\percent}$, we only focus on those NV centers containing this isotope. Hence, the NV spin Hamiltonian with the electron and nuclear spin matrices $\hat{S}$ and $\hat{I}$ for an applied magnetic field reads as\cite{charnock2001combined}

\begin{equation}
	\mathcal{H} = D\hat{S}^2_z + g_e\mu_B B \hat{S}_z + A\hat{\underline{S}}\,\hat{\underline{I}} + Q\hat{I}^2_z + g_n\mu_B B \hat{I}_z,
\end{equation}

where the first term with $D = \SI{2870}{MHz}$ denotes the zero field splitting of the $\left|0\right>\rightarrow\left|\pm1\right>$ transition and the second term with the electron g-factor $g_e$ corresponds to the Zeeman energy with $\mu_B$ the Bohr magneton. The other three terms represent the nuclear spin interacting with the electron spin (hyperfine splitting tensor $A$), the nuclear quadrupole interaction $Q$ and the nuclear Zeeman energy $g_n \mu_B$. At the applied magnitude of the magnetic field of $\left|B\right|\sim\SI{230}{Gauss}$, the interaction with the $I = 1$ nuclear spin leads to a splitting of the given energies into triplets\cite{doherty2012theory} of the $\left|0\right>\rightarrow\left|\pm1\right>$ levels respectively (see fig.\ref{fig:pulsedODMR_meas}).\\

\begin{figure}
	\begin{subfigure}{.5\textwidth}
		\centering
		\includegraphics[width=.8\linewidth]{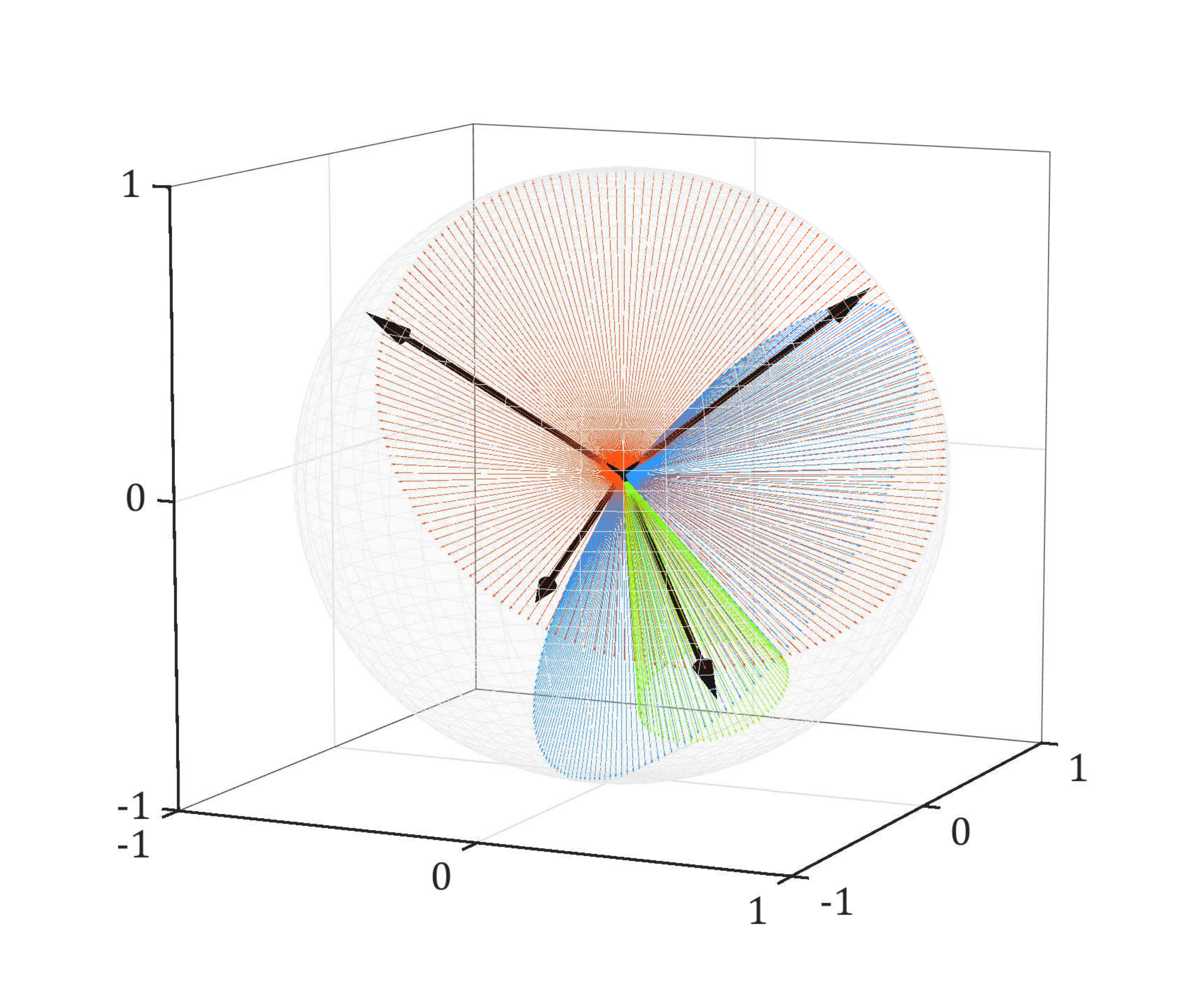}
		\label{subfig:Three_Cone_Model}
		\caption{}
	\end{subfigure}%
	\begin{subfigure}{.5\textwidth}
		\centering
		\includegraphics[width=.8\linewidth]{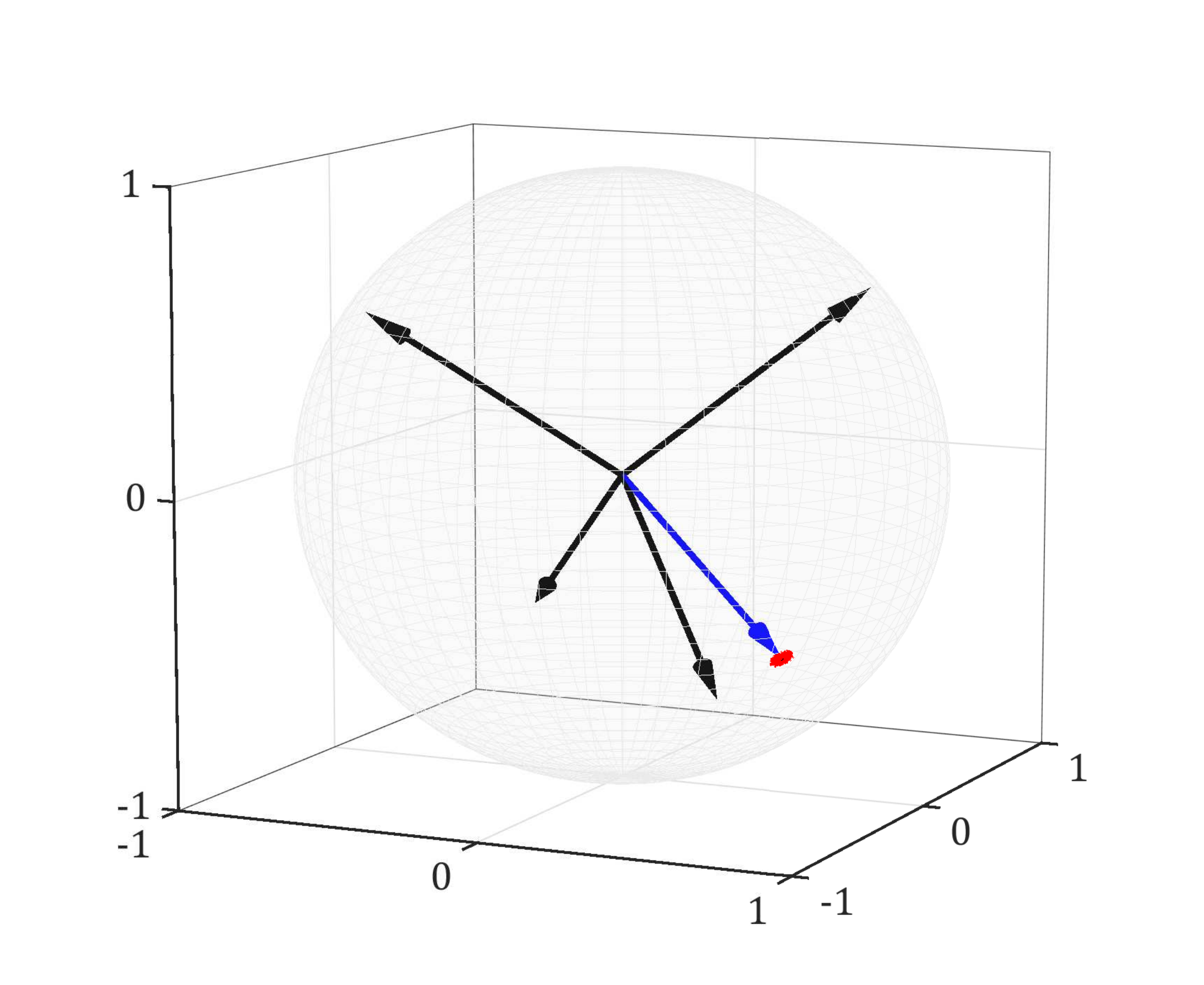}
		\label{subfig:B_Vector_Solution_Sphere}
		\caption{}
	\end{subfigure}
	\caption{(a) Exemplary plot of the intersecting cone model (intersection of three magnetic field vector cones). The three cones for $[1\bar{1}\bar{1}]$(green), $[\bar{1}1\bar{1}]$(orange), and $[\bar{1}\bar{1}1]$(blue) are inserted exemplarily.(b) Result for the reconstruction of the B-field vector(blue) with the help of the three measured NV center axes shown in subfigure a. For this simulation, we used the axes $[1\bar{1}\bar{1}]$, $[\bar{1}1\bar{1}]$, and $[\bar{1}\bar{1}1]$ and the red area represents the angular covariance distribution of the result.}
	\label{fig:NV_schematic3D}
\end{figure}

In order to determine the B-field orientation, either two or three different tilt angles for the same magnetic field, but different NV axes, need to be quantified. Hence, we keep the magnetic field constant and measure the B-field tilt angle $\theta_{\mathrm{i}}$ for different NV axes, with $i \in [1,2,3,4]$ denoting the four possible NV orientations. The spectra for the different NV orientations are measured using a pulsed Optically Detected Magnetic Resonance (ODMR)\cite{jelezko2004read} scheme in which a  $\frac{\pi}{2}$-pulse is performed before each readout instead of a continuous wave manipulation (see fig.\ref{fig:pulsedODMR_meas}). Additionally to the small linewidth and high accuracy of the ODMR signal, this measurement allows to resolve the hyperfine interaction between the NV-spin state and the nitrogen nucleus and for all further calculations, we used the $m_{I} = 0$ transition exclusively.\\
We calculated the magnetic field miss-alignment angle $\theta$, utilizing the $\left|{0}\right>\rightarrow\left|{-1}\right>$ and $\left|{0}\right>\rightarrow\left|{+1}\right>$ transitions symmetry with respect to the zero field splitting\cite{balasubramanian2008nanoscale}. However, for an arbitrarily oriented magnetic field $\pmb{B}$, only the tilt angle $\theta$ can be determined in this way. Thus, the possible orientations of the magnetic field vector are given by a cone around the NV axis defined by the opening angle $\theta$, leaving infinite possible orientations available (see fig.\ref{fig:NV_schematic3D}a). In the following, the relative tilt angle is going to be denoted as $\theta$, whereas the angles of the three dimensional spherical coordinates are called $\psi$ and $\phi$.\\
For the signal normalization, we included $10$ Rabi cycles and taking the Rabi states as maximal bright and dark signal, hence, taking the difference as full contrast.\\ 
If only two of the four different NV axes are measured, there exist two possible intersections of the B-field cones. This allows two equally likely B-field orientations symmetrically with respect to the NV axes (except for the special case of $\theta_{\mathrm{i}} + \theta_{\mathrm{j}} = \SI{109.47}{\degree}$, which gives exactly one solution in the plane spanned by those two NVs).\\
In the case of three measured NV-orientations, the B-field orientation can be determined precisely. The result of one exemplarily calculated B-field vector is illustrated in fig.\ref{fig:NV_schematic3D}b.

\begin{figure}
	\begin{subfigure}{.5\textwidth}
		\centering
		\includegraphics[width=.8\linewidth]{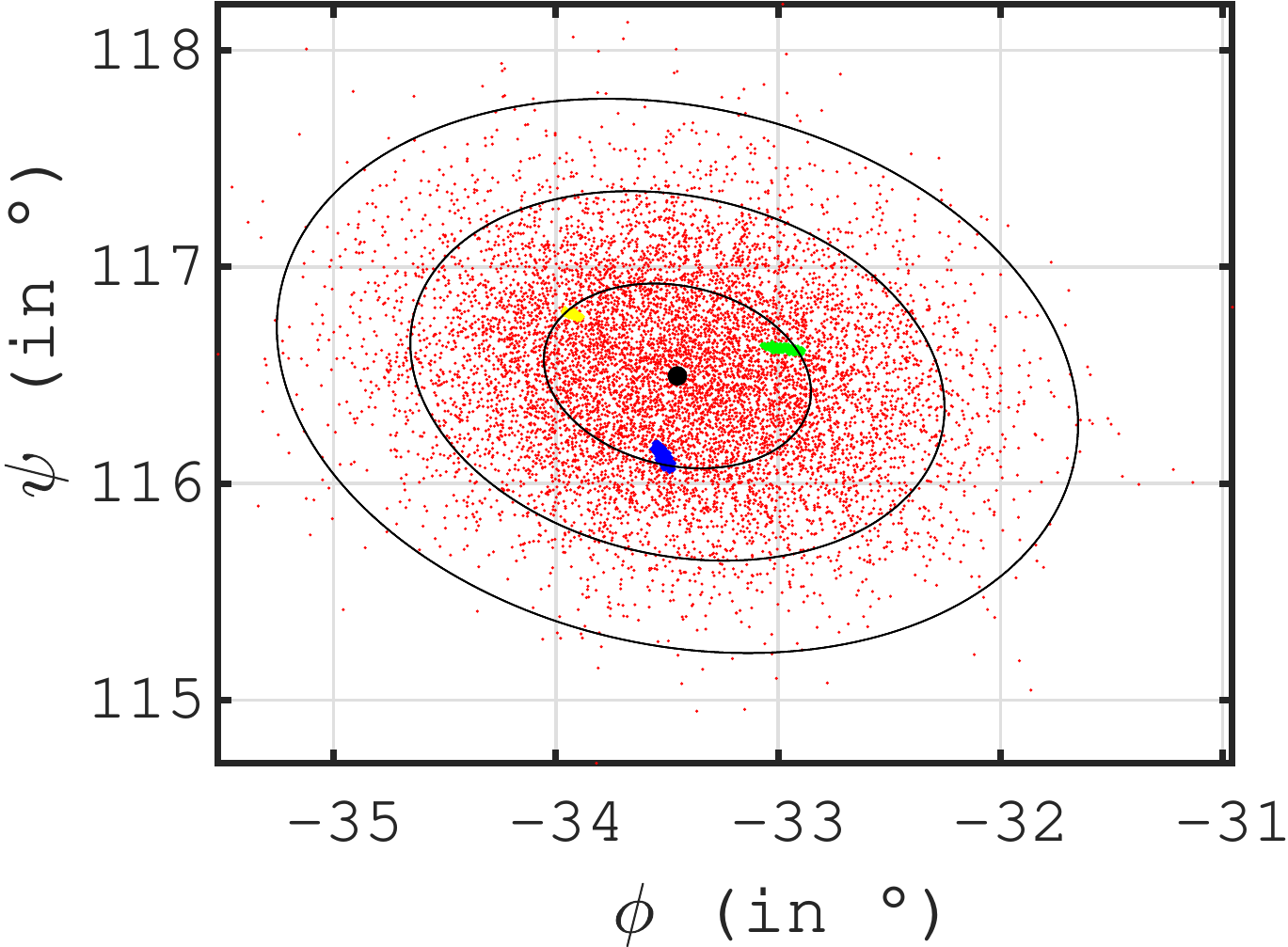}
		\label{subfig:Solution_Angle_Distribution123}
		\caption{}
	\end{subfigure}%
	\begin{subfigure}{.5\textwidth}
		\centering
		\includegraphics[width=.8\linewidth]{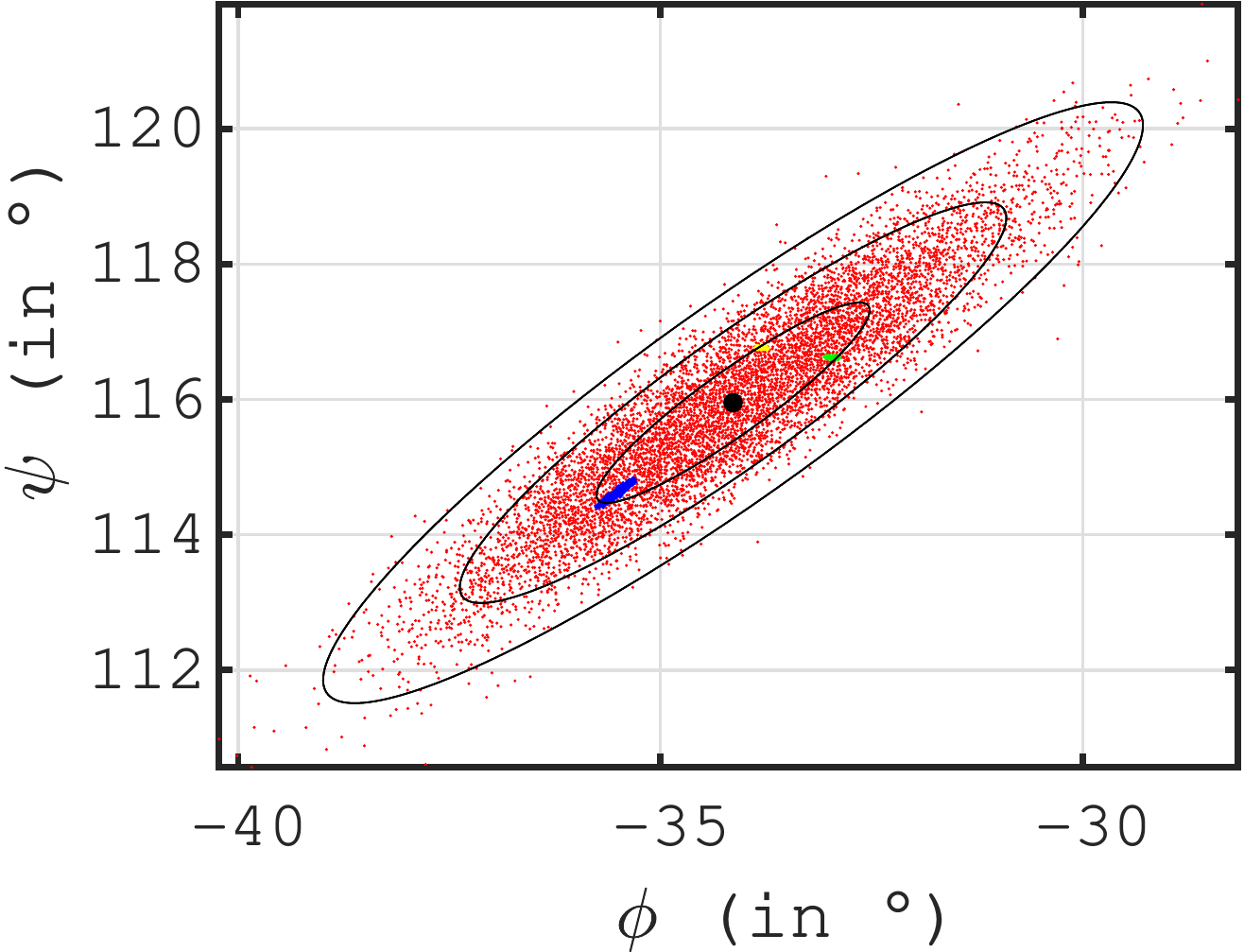}
		\label{subfig:Solution_Angle_Distribution124}
		\caption{}
	\end{subfigure}
	\caption{Angular distribution ($\phi\;\text{and}\;\psi$) of the resulting B-field orientation vector. The covariance matrix is calculated as the smallest enclosing ellipse of the three subsets. The black circles represent the standard deviation intervals $\sigma$, $2\sigma$, and $3\sigma$. Simulation results for NV axes (a) 123 (the combination of NV1, NV2, and NV3) and (b) 124 (the combination of NV1, NV2, and NV4) with NV1:$([1\bar{1}\bar{1}],\theta_{1} = \SI{13.44\pm0.15}{\degree})$, NV2:$([\bar{1}1\bar{1}], \theta_{2} = \SI{62.924\pm0.005}{\degree})$, NV3:$([\bar{1}\bar{1}1],\theta_{3} = \SI{66.612\pm0.006}{\degree})$,and NV4:$([111],\theta_{4} = \SI{83.21\pm0.13}{\degree} )$. The three distributions (green, yellow, and blue) represent the individual simulation sets. The final covariance distribution is depicted in red.}
	\label{fig:Solution_Angle_Distribution}
\end{figure}

For the calculation, we normalize the NV axes (NV1: $[1\bar{1}\bar{1}]$, NV2: $[\bar{1}1\bar{1}]$, NV3: $[\bar{1}\bar{1}1]$, and NV4: $[111]$) and contract them into the origin, drawing a unit sphere around it which contains the four normalized NV axes(\ref{fig:NV_schematic3D}a black vectors and grey sphere). As shown in fig. \ref{fig:NV_schematic3D}a, the insertion of the possible magnetic field vectors into the NV center conformation, leads to the intersection of the individual cones. Due to the symmetric interaction of the B-field with the NV center axes in their tetrahedral conformation, the parallel and the anti-parallel orientation of the B-field vector lead to the same outcome of the measurement and calculation of $\theta_{\mathrm{i}}$.\\
 The cut of this unit sphere with the B-field cones lead to plane equations described by

\begin{equation}
	\mathrm{sgn}(NV_{i,x})B_x + \mathrm{sgn}(NV_{i,y})B_y + \mathrm{sgn}(NV_{i,z})B_z = \sqrt{3} \cos\left( \theta_{\mathrm{i}} \right),
	\label{eq:cone_sphere_plane}
\end{equation}

with the signum function giving only the sign of the i-th NV axis direction vector component and $\left(B_x,B_y,B_z\right)^T$ the B-field vector solution candidate. Note that geometrically those cones do not intersect if $\theta_i+\theta_j < \SI{109.47}{\degree}$. Therefore, if $\theta_{\mathrm{i}} > \SI{54.735}{\degree}$, one needs to do a point mirroring with respect to the origin and define the new angle as $\theta_i' = \SI{180}{\degree} - \theta$. For every B-field orientation, there exists one NV axis fulfilling $\theta_{\mathrm{j}} < \SI{54.735}{\degree}$, whereas the others show $\theta_{\mathrm{i}} > \SI{54.735}{\degree}$ for $i \neq j$. There is only one relative B-field orientation for which $\theta_{\mathrm{i}}\equiv\SI{54.735}{\degree}$ holds true for all four $i$.\\
Additionally to the solution in eq.\ref{eq:cone_sphere_plane}, the B-vector $\pmb{B} = \left(B_x,B_y,B_z\right)^T$ for the solution of linear equations needs to point onto the surface of the unit sphere, hence, its length $|\pmb{B}| = 1$ has to be fulfilled as well. Using twice eq.\ref{eq:cone_sphere_plane} (e.g. NV1 with $\theta_1$ and NV2 with $\theta_2$) together with the normalization of the B-field gives us three linear equations for the subset of the three individual measured B-field tilt to NV direction pairs. As a consequence, we obtain three systems of three linear equations for the two axes subsets $i,j$, with the solutions depicted as yellow (NV1 with NV2), green (NV1 with NV3), and blue (NV2 with NV3) distributions in fig.\ref{fig:Solution_Angle_Distribution}. Those distributions are obtained by taking one million samples from each normally distributed angles $\theta_i$ independently and solve the set of linear equations to calculate the cone overlap respectively. The expectation value $\pmb{\mu}$ (black middle in fig.\ref{fig:Solution_Angle_Distribution}) and the standard deviation $\pmb{\sigma}$ are hereby defined as the center and covariance matrix of the smallest enclosing ellipse containing $\SI{99.7}{\percent}$ of all simulated points.

In the case of an error free measurement and calculation of the cone angles $\theta_i$, it would be possible to solve the set of equations for the three axes at once, but since this cannot be granted, it is important to consider each intersect individually.\\ 
If the linear equation system is solved for the three plane equations (\ref{eq:cone_sphere_plane}) at once, the individual error is neglected, for instance, if one measurement was defective, this error would get lost in the averaging of the solver instead of being emphasized by the distortion of a covariant distribution (see fig.\ref{fig:Solution_Angle_Distribution}b).\\
Since we are only interested in the directional vector of the magnetic field, the radius is set to one ($r\equiv 1$). Hence, also the covariance matrix is only angle dependent, which leads to a bivariate normal distribution with $\pmb{B},\pmb{x} \in \mathcal{R}^2$ and $\pmb{\Sigma} \in \mathcal{R}^{2x2}$ and the probability density function

\begin{equation}
	p(\pmb{x};\pmb{B},\pmb{\Sigma}) = \frac{1}{2\pi\left|\pmb{\Sigma}\right|^{1/2}} \exp\left\{ -\frac{1}{2}(\pmb{x}-\pmb{B})^T\pmb{\Sigma}^{-1}(\pmb{x}-\pmb{B}) \right\}.
\end{equation}

If we apply the described simulations to the measurement data given in fig. \ref{fig:pulsedODMR_meas} with once the NV axes $123$ (the combination of NV1, NV2, and NV3) and secondly the NV axes $124$ (the combination of NV1, NV2, and NV4), this results in the expectation value vectors

\begin{equation}
	\pmb{B}_{123}\left(\phi,\psi\right) = \begin{pmatrix} \SI{-33.45}{\degree}\\\SI{116.50}{\degree}\end{pmatrix}
	\quad\text{and}\quad 
	\pmb{B}_{124}\left(\phi,\psi\right) = \begin{pmatrix} \SI{-34.14}{\degree}\\\SI{115.95}{\degree}\end{pmatrix}
\end{equation}
with the corresponding covariance matrices
\begin{equation}
\pmb{\Sigma}_{123} =
	\begin{pmatrix}
		\SI{0.36}{\degree} & \SI{-0.05}{\degree}\\
		\SI{-0.05}{\degree} & \SI{0.17}{\degree}
	\end{pmatrix}
	\quad\text{and}\quad
	\pmb{\Sigma}_{124} =
	\begin{pmatrix}
		\SI{2.6}{\degree} & \SI{1.7}{\degree}\\
		\SI{1.7}{\degree} & \SI{1.2}{\degree}
	\end{pmatrix}.
	\label{eq:covariance_matrix}
\end{equation}

The covariance matrices show that for well determined B-field orientations, the overall error interval is below $\SI{0.4}{\degree}$. For the fourth axes with a misalignment of $\theta_{4} = \SI{83.21\pm0.13}{\degree}$, the ODMR measurement showed a washed out signal of the three transition frequencies (see Supporting Information fig. 1), whereby the assignment of the proper $\left|{m_s = 0, m_I = 0}\right>\rightarrow\left|{m_s = \pm1, m_I=0}\right>$ frequencies is quite difficult. This is visible in the increase of the individual covariances given in $\pmb{\Sigma}_{124}$.

\section{Discussion}
	In this paper, we have proposed a technique and experimental realization of single spin magnetometry to determine the orientation of a $3D$ static magnetic field by utilizing single NV defects in diamonds.\\
	Regarding the measurement duration, the reconstruction of a B-field vector mainly depends on the shot noise limited pulsed ODMR measurement of the $\left|{0}\right>\rightarrow\left|{-1}\right>$ and $\left|{0}\right>\rightarrow\left|{+1}\right>$ transitions for the three different NV orientations. This leads for the presented magnetic field reconstruction to a typical duration of roughly one hour.\\
	When determining the B-field orientation by common pulsed ODMR measurements and current technology, we obtain the presented results with a small error of less than $\SI{0.4}{\degree}$. However, in comparison to the initial error of the cone angles $\theta_i$, the error interval obtained from the simulation is more than one order of magnitude larger. Therefore, the main contribution to the error interval for this reconstruction does not originate in the linewidth of the pulsed ODMR measurement, but results due to the deviation in the overlap of the three cones, evident in fig. \ref{fig:Solution_Angle_Distribution}. This deviation can arise from strain, not taken into account for the calculation of the cone angles $\theta_i$. Nevertheless, for an arbitrary magnetic field, the B-field vector solutions given by the separate cones can point toward almost all possible orientations on the sphere. In contrast, the here presented technique allows to narrow this multitude of orientations down to only a very small fraction equivalent to a single axis miss-alignment on the order of $\SI{0.4}{\degree}$.\\
	In summary, the sensitivity for changes of the magnetic field direction as well as the overall precision are in principle limited only by the linewidth of the individual ODMR spectra. Hence, if strain is accounted for, a lower bound is determined by the lifetime of the utilized NV centers.

	Examples of full B-field determination in the past rely on an NV ensemble where NV centers with all the four different orientations in the $[111]$ crystal direction are in close proximity\cite{maertz2010vector}. Another approach uses ensemble measurements with continuous wave readout and frequency multiplexing to determine the ODMR spectra\cite{schloss2018simultaneous}, however it is conceivable that this approach could be extended to the single site level. Moreover, a strongly coupled carbon nuclear spin in close proximity can be used for B-field determination \cite{jiang2018estimation}. In contrast, the here presented approach uses single NVs measured in the same confocal volume without any further prerequisite. Due to the simplicity of the setup, our technique is generally applicable and thus can be used with most setups without any further technical modification. Additionally, the lack of specific attributes for the used NV centers allows to employ this scheme for all samples and not only specific sites, in particular all necessary measurements can be performed on each sample and sample exchange is not required. This gives us the possibility to characterize the magnetic field at the focus of the microscope sample universally. Therefore, one can even use this measurement as calibration for other non NV based measurements.\\

	One further important application of this technique is the determination of the absolute NV-center axis orientation in the diamond lattice. Thus allowing to obtain atomic structure information by the use of macroscopic measurements. To achieve this, a second objective has to be implemented atop of the diamond sample (Supporting Information fig. 2). This objective replaces the magnets in the original setup (fig. \ref{fig:Confocal_Head_NV_Lattice}a), and has to be mounted in a tilted manor with respect to the optical axis. With a separate excitation of the NV center via this objective, the indistinguishability of the two different axes of one polarization can be resolved by comparison of the individual fluorescence signals.

\begin{acknowledgement}
The author thanks Christian Osterkamp and Johannes Lang (Institute for Quantum Optics, Universit\"at Ulm) for the overgrowth and 
nitrogen vacancy implantation of the diamond sample.
This work was funded by the center for Integrated Quantum Science and Technology (IQST) and by the DFG, Sonderforschungsbereich 1279.

\end{acknowledgement}

\begin{suppinfo}

The following files are available free of charge.
\begin{itemize}
  \item Supporting Information: Experimental and technical information 
\end{itemize}

\end{suppinfo}

\providecommand{\latin}[1]{#1}
\makeatletter
\providecommand{\doi}
  {\begingroup\let\do\@makeother\dospecials
  \catcode`\{=1 \catcode`\}=2 \doi@aux}
\providecommand{\doi@aux}[1]{\endgroup\texttt{#1}}
\makeatother
\providecommand*\mcitethebibliography{\thebibliography}
\csname @ifundefined\endcsname{endmcitethebibliography}
  {\let\endmcitethebibliography\endthebibliography}{}

\end{document}


\maketitle
	\subsection*{Diamond Sample}
		The diamond sample we used for this work is a single-crystal electronic grade diamond (Element Six).  It contains substitutional nitrogen and boron atoms in concentrations less than 5 parts per billion
(ppb) and 1 ppb respectively. Prior to the nitrogen implantation, the diamond was laser cut and polished with (100) surface to a height of $\SI{35}{\micro\meter}$ (Applied Diamond Inc.). The nitrogen implantation was done with an implantation energy of $\SI{5}{keV}$ into a beforehand via chemical vapor deposition (CVD) grown $^{12}C$ enriched layer. These processes were done by members of the Quantum Optics department of Ulm University.

	\subsection*{Optical Setup}
		We used a home build confocal microscope to detect and manipulate the single NV centers. For the excitation, a $\SI{519}{\nano\meter}$ laser diode system from TOPTICA Photonics (iBeam-Smart-PO) with pulse option was used. The diamond sample was placed on a $\sim\SI{100}{\micro\meter}$ cover slide and measured through the cover slide, as well as through the diamond with a Nikon CFI A-Apo 100x oil objective (NA = 1.45). To move the sample, a piezoelectrical scanner from Physical Instruments (PInano P-545) was used. A single photon counting module from Excelitas Technologies (SPCM-AQRH-14) was utilized to detect the emitted NV fluorescence. To cut off the excitation wavelength, we placed a $\SI{635}{\nano\meter}$ long-pass filter in front of the photo diode. Gated photon counting is realized by using an FPGA developed and provided by the Quantum Optics department of Ulm University. 
	
	\subsection*{Microwave Setup}
		To generate the microwave pulses, an arbitrary waveform generated from Tektronix (AWG 70001A) in combination with a microwave amplifier (AR-50HM1G6AB-47) from Amplifier Research are used. A copper wire (diameter $d=\SI{25}{\micro\meter}$) spanned over the diamond surface serves as antenna. Typical distances between measured NV centers and the antenna is on the order of $\SI{15}{\micro\meter}$.
		
	\subsection*{Magnet Setup}
		To provide a homogenous magnetic field, we used several cubic neodymium magnets, which were arranged in a cross-like shape. The magnet position was controlled by three linear translation stages (LS-110) and one rotation stage (PRS-110) from Physical Instruments (PI miCos). Hence, a repeatability down to $\SI{0.5}{\micro\meter}$ in magnet positioning was guaranteed.
		
	\subsection*{Software}
		For control of the setup components, generation of the microwave sequences, timing of the readout, and analysis of the signal, the modular python suite Qudi\cite{binder2017qudi} was employed. The simulation and reconstruction of the magnetic field vector was done in MathWorks MATLAB (Version R2019a).
	
	\subsection*{Measurement Techniques}
		The employed pulsed ODMR technique\cite{jelezko2004read} consisted of a single $\pi$-pulse with varying frequency and a following gated laser pulse of $\SI{3}{\micro\second}$. In all measurements, the applied Rabi frequency for the pulses was on the order of $\SI{250}{kHz}$. To obtain the state information of the NV center, we analyze the first $\SI{300}{\nano\second}$ of the detected signal.
		
	\subsection*{Additional Graphics}
	
		\begin{figure}[hp]
			\includegraphics[width=.5\linewidth]{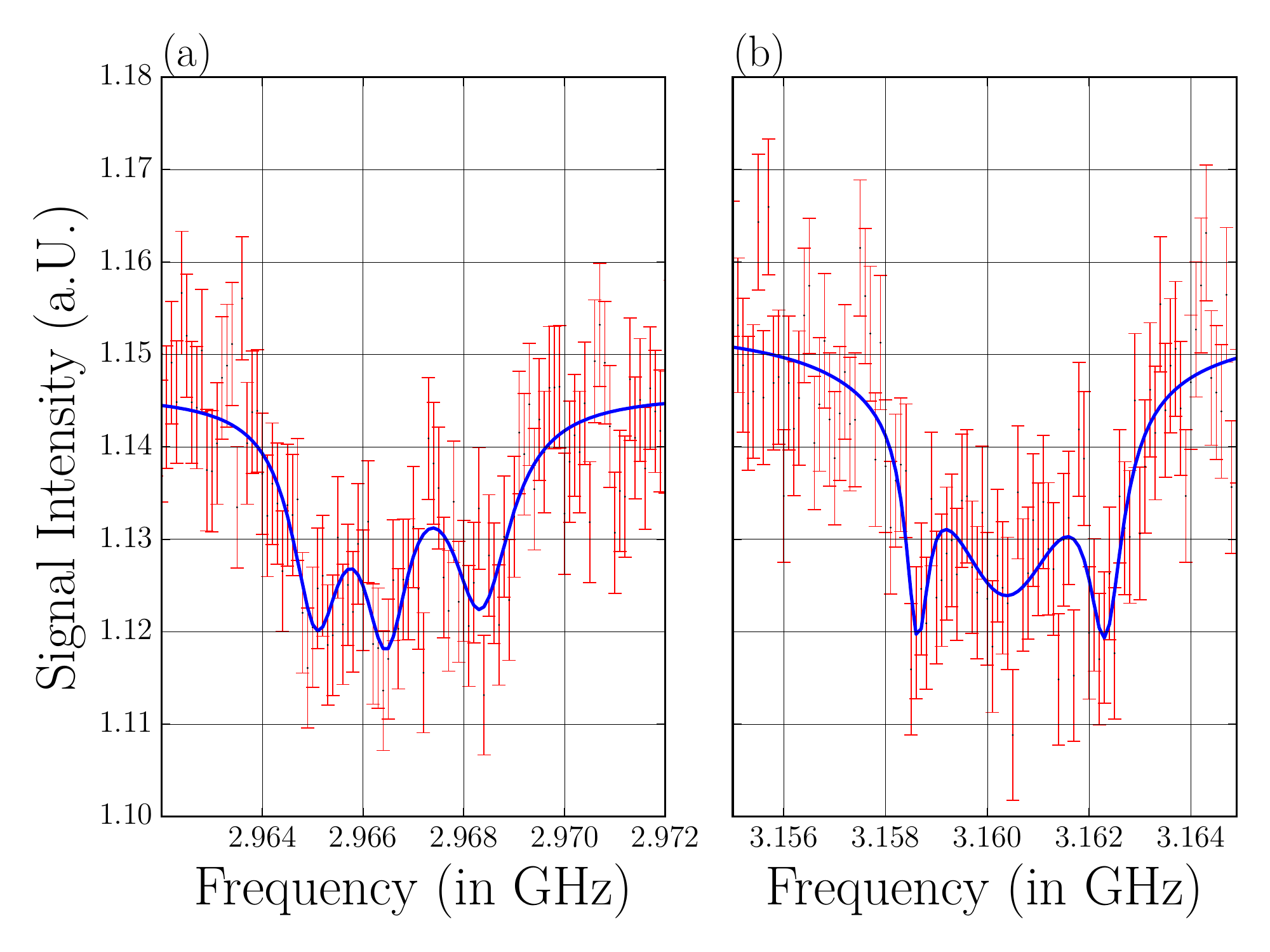}
			\centering
			\caption{Results for the pulsed ODMR measurement normalized to the maximum Rabi contrast. Measurements for the fourth NV orientation with respect to the B-field are shown. The transition frequencies $\left|0\right>\rightarrow\left|-1\right>$ (a) and $\left|0\right>\rightarrow\left|\pm1\right>$ (b) with resulting angle $\theta_{\mathrm{4}} = \SI{83.21}{\degree}$. Due to the strong misalignment, the transition frequencies smear out. Hereby, the Rabi frequency for the $\pi$-pulses of the measurements was $\SI{500}{kHz}$.}		
			\label{fig:Polarization_Axes}
		\end{figure}
		
		\begin{figure}[hp]
			\includegraphics[width=.5\linewidth]{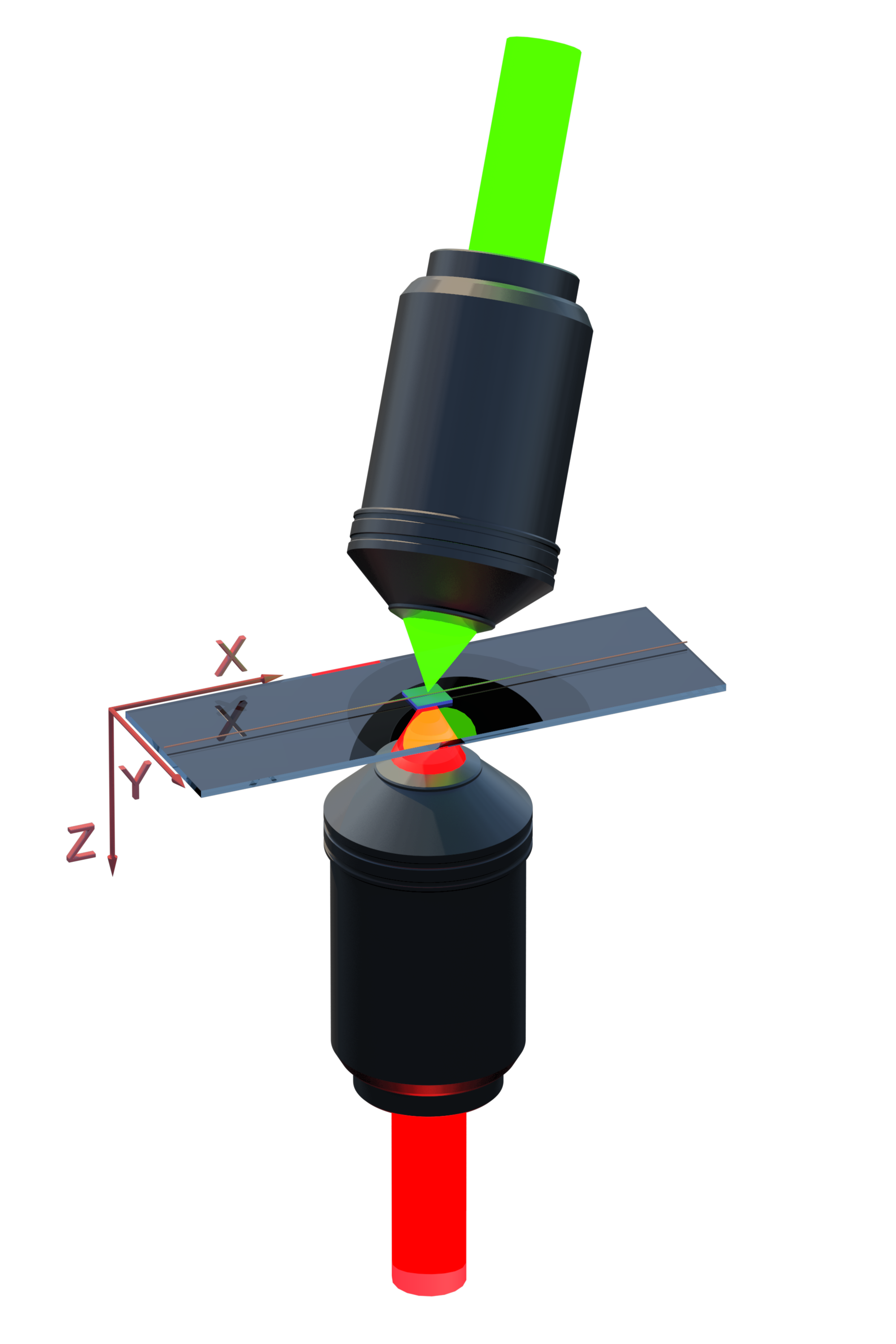}
			\centering
			\caption{Schematic, illustrating the alternative experimental geometry with both objectives focused onto the same spot. The excitation is done via the second objective (green optical pathway), which replaces the magnets, while the detection is still performed with the lower objective (red optical pathway).}		
			\label{fig:Setup_Two_Objectives}
		\end{figure}